\documentstyle[12pt]{article}
\normalbaselines
\begin{document}
\begin{center}
{\large \bf
NEW SYMMETRIES IN MATHEMATICAL PHYSICS EQUATIONS}\\[2mm]

G.A. Kotel'nikov\\{\it Russian Research Center "Kurchatov Institute",
123182 Moscow, Russia}\\
(e-mail: kga@kga.kiae.su)\\[5mm]
\end{center} 

\begin{abstract}
\smallskip

    An algorithm for studing the symmetrical properties of the partial 
differential equation of the type $L\phi(x)=0$ is proposed. By symmetry
of this equation we mean the operators $Q$ satisfying commutational
relations of order $p$ more than $p=1$ on the solutions $\phi(x)$:
$[L \ldots[L,Q] \ldots]\phi(x)=0$. It is shown, that within the
framework of the proposed method with $p=2$ the relativistic D'Alembert
and Maxwell equations are the Galilei symmetrical ones. Analogously, with
$p=2$ the Galilei symmetrical Schr\"odinger equation is the relativistic
symmetrical one. In both cases the standard symmetries are realized with
$p=1$.

\end{abstract}

\section{Introduction}
The symmetry properties of mathematical physics
equations contain the important information about objects of research.
\par Some receptions are offered for symmetry research:
the classical and  the modified Lie
methods \cite {Ovs78}, \cite {Fus90}, the non-Lie ones 
\cite{Fus90}, \cite{Zhe86}, the renormgroup concept 
\cite {Shi94}, the method for search of the conditional symmetries 
\cite{Vor86},\cite{Olv86},
the theoretical-algebraic approach \cite{Mal79}.
The purpose of the present
work  is the formulation of the new method
for obtaining the additional  information about symmetrical
properies of equations.
For  research
we choose the D'Alembert, Maxwell and Schr\"odinger equations.
\section{Method of Research}
\par We begin with a definition of symmetry, which we shall name as
extended one.
\par Let the equation be given in the space $R^n(x)$
\begin{equation}
\label{f1}
L\phi(x)=0{,}
\end{equation}
where $L$ is a linear  operator.
\newtheorem{definition}{Definition}
\begin{sloppypar}
\begin{definition} By the symmetry of  Eq. (1) we shall
mean a set of operators $\{Q^{(p)}\}$, $p=1,2,\ldots,n,\ldots$,
if the result of the successive $(p-1)$-fold
action of the operator $L$ on an operator $Q^{(p)}$ transforms a nonzero
solution  $\phi(x)$ into another
solution $\phi'(x)=L^{(p-1)}Q^{(p)} \phi(x)\not=0$.
\end{definition}
\end{sloppypar}
    From Definition follows, that operators $Q^{(p)}$ satisfy commutational 
relations of order $p$
\begin{equation}
\label{f2}
[L\ [L\ldots[L,Q^{(p)}] \ldots]\ ]_{(p-fold)}\phi(x)=0.
\end{equation}

    The extended Definition:
\begin{itemize}
\begin{sloppypar}
\item includes the understanding of symmetry  in the case when 
Eq. (\ref{f2}) is fulfiled on a set of arbitrary functions,
that is equivalent to 
$[L\ [L\ldots[ L, Q^{(p)} ]\ldots]]_{(p-fold)}=0$ \cite {Kot83};
\end{sloppypar}
\item contains the standard understanding of symmetry, when \
$[L,Q^{(1)}] \phi(x)=0$ \cite {Mal79};
\item  includes the understanding of symmetry
in quantum mechanics sense $[L,Q^{(1)}]=0$;
\item differs  from the standard one,
as    in  the framework  of  the latter by the operators of symmetry
we   should mean   not  the  operators  $\{Q^{(p)}\}$,  but the
operators $\{X^{(1)}=L^{(p-1)}Q^{(p)}\}$ \cite {Lez86}.
\end{itemize}
The question is how practically to find them. In the present work it
is  decided  by  analogy with the modifed Lie algorithm \cite {Fus90}.
Below we consider the case, when $p = 2$.
\par Let us introduce a set of operators
\begin{equation}
\label{f3}
Q^{(1)}=\xi_1^a(x)\partial_a+\eta_1(x){;}
\end{equation}
\begin{equation}
\label{f4}
Q^{(2)}=\xi_2^a(x)\partial_a+\eta_2(x){,}
\end{equation}
which have the following commutation properties
\begin{equation}
\label{f5}
[L,Q^{(1)}]=\zeta_1(x)L{;}
\end{equation}
\begin{equation}
\label{f6}
[L\ [L,Q^{(2)}]\ ]=\zeta_2(x)L.
\end{equation}
The given expresions are operator's versions of  the
extended Definition of symmetry. Here $\partial_a=\partial/\partial{x^a}$;
$a=0,1,\ldots{n-1}$; $\xi^a(x)$, $\eta(x)$, $\zeta(x)$ \ are unknown functions;
the summation is carried out over a twice repeating index;
the unknown functions may be found by  equaling the coefficients at
identical derivatives in the left and in the right parts of ratios
and  by integrating  the  set of determining 
differential equations available.
\par     After integrating the general form of the operators $Q$ may be
recorded as a linear combination of the basic elements 
$Q_\alpha^{(1)}$ and $Q_\mu^{(2)}$, on which, by analogy with
\cite{Fus90} we impose the condition  
to  belong in  Lie algebras:
\begin{equation}
\label{f7}
A^1:[Q_\alpha^{(1)},Q_\beta^{(1)}]=C_{\alpha\beta\gamma}Q_\gamma^{(1)}{;}
\end{equation}
\begin{equation}
\label{f8}
\begin{array}{c}
A^2:[Q_ \epsilon^{(1)},Q_ \delta^{(2)}]=C_{\epsilon\delta\chi}Q_\chi^{(1,2)}; \\
\hspace{+8mm} \lbrack Q_ \mu^{(2)},Q_ \nu^{(2)} \rbrack =
C_{\mu\nu\sigma} Q_ \sigma^{(1,2)}.
\end{array}
\end{equation}
Here \  $C_{\alpha\beta\gamma}$, $C_{\mu\nu\sigma}$ \ are the structural
constants;
operators $Q_\chi^{(1,2)}$, $Q_\sigma^{(1,2)}$
belong to the \ {sets} of operators $\{Q^{(1)}, Q^{(2)}\}$.
\par   With help of
the  Lie equations we transfer from Lie algebras to the Lie groups
\begin{equation}
\label{f9}
dx^a{'}/d\theta=\xi^a{(x')}{,}
\end{equation}
where $x^a{'}_{(\theta=0)}=x^a$; \ $a=0,1,\ldots{n-1}$; \ $\theta$ \ is a
group parameter \cite{Fus90}.
\begin{sloppypar}
    For the law of field transformation to be found, instead of
integrating the Lie equations
$d\phi'{(x')}/d\theta=\eta{(x')}\phi'{(x')}$, \
$\phi'{(x')}_{(\theta=0)}=\phi{(x)}$ \cite{Fus90}  we shall take
the reception \cite{Kot86}, which we shall illustrate
by  example of one-component field.
\end{sloppypar}
Let us introduce such a weight function $\Phi(x)$ in the field 
transformation law, that
\begin{equation}
\label{f10}
\phi'(x')=\Phi(x)\phi(x).
\end{equation}                                              
We choose the function $\Phi(x)$  so that  Eq. (\ref{f1}) should
transform into itself in accordance with the generalized understanding of
symmetry 
because of the following additional condition (the set of engaging
equations)
\begin{equation}
\label{f12}
A\Phi(x)\phi(x)=0, \ L\phi(x)=0.
\end{equation}
The former is obtained by replacing  the variables in the initial
equation $L'\phi'(x')=0$. Formula (\ref{f10}) corresponds to 
a linearization of the transformed unprimed equation at replacing $x'=x'(x)$,
$\phi'(x')=\phi'(\Phi (x) \phi (x))$ \cite{Kot86}.
If here $A=L$, we shall call the symmetry the classical symmetry
and if $A\not=L$, we shall refer it to as the generalized one.
By solving Set (\ref{f12}),
the weight function $\Phi(x)$ can be put in
conformity to  each field function $\phi(x)$ for ensuring
the transition $L' \to L$.
\begin{sloppypar}
    Instead of solving Set (\ref{f12}), the weight  function
may be found on  the base of the symmetry approach.  As far
as $\phi'(x')=Q^{(1)'} \phi(x')$ is a solution too, and
$\phi'(x')= \Phi(x)\phi(x)$ we have \cite{Kot86}
\end{sloppypar}
\begin{equation}
\label{f13}
\Phi(x)={\phi'(x'\to x)\over\phi(x)}\in\{{\phi(x'\to x)\over\phi(x)};
{1\over\phi(x)};
{Q_\alpha^{(1)'}\phi(x'\to x)\over\phi(x)}; 
{[L',Q_\mu^{(2)'}]
\phi(x'\to x)\over
\phi(x)};\ldots\}.
\end{equation}
Here the dots correspond to a consecutive action of the
operators $Q_\alpha^{(1)'}$ and $[L',Q_\mu^{(2)'}]$ on a solution $\phi(x')$.
Thus, for the function $\Phi(x)$ to be found  it is necessary to turn to the 
unprimed variables  in the primed
solution $\phi'(x')$, and to divide 
the result available  by the unprimed  solution  $\phi(x)$ \cite{Kot86}.
\par    After  finding the  weight  functions $\Phi(x)$ the task about the
symmetry of  Eq. (\ref{f1}) for one-component field may be
thought  as completed in the definite sense, namely:  the set of the
operators of symmetry and the corresponding Lie  lgebras are indicated
for $p=1$  and  $p=2$;  the groups of symmetry are
restored by the given algebras;  with help of the weight functions the
transformational properties of field $\phi(x)$  are determined.
\par   The proposed method allows generalization to the case of 
multicomponent field and a symmetry of order more high, than $p=2$.
\section{Application of the Method}
\subsection{The Galilei symmetry of D'Alembert equation, p=2}
\begin{equation}
\label{f14}
L_D \phi(x)=\Box \phi(x)=(\partial_{tt}/c^2 - \triangle)\phi(x)=0
\end{equation}
\begin{equation}
\label{f15}
\phi(x)=exp(-ik.x)=\omega (t - {\bf n.x }/c)
\end{equation}
Generator of space-time transformations and its commutational properties:
\begin{equation}
\label{f16}
H_1=x_0\partial_1; \ [\Box [\Box,H_1]=0.
\end{equation}
Conditions of transfere of equation  (\ref{f14}) into itself:
\begin{equation}
\label{f17}
[(\partial_0 + \beta \partial_1 )^2 /\lambda^ 2 - \bigtriangleup ]
\Phi_D (x) \phi (x) =0; \ \Box \phi (x) =0.
\end{equation}
Weight function:
\begin{equation}
\label{f18}
\Phi_D (x)=
exp\{-{i\over \lambda}[(1-\lambda )k.x - \beta \omega (n_x t - x/c)]\}.
\end{equation}
Transformational properties of solution (\ref{f15}):
\begin{equation}
\label{f19}
exp(-ik'.x')=
exp\{-{i\over \lambda}[(1-\lambda )k.x - \beta \omega (n_x t - x/c)]\} 
exp(-ik.x).
\end{equation}
\par Here $x_0=x^0=ct$, $t$ is the time;
$x_k=-x^k$, $k=1,2,3$, $x^{1,2,3}=x,y,z$ are the space variables; 
$\omega$ is the frequency.
\subsection{The Lorentz symmetry of Schr\"odinger equation, p=2}
Let us investigate equation, which we name  relativistic Schr\"odinger
equation, and next  transfer to the known non-relativistic one. We have:
\begin{equation}
\label{f20}
L_S \psi{(x)}=(i\hbar\partial_t+{\hbar^2\sqrt{1-\beta^2}\over2{m_0}}\triangle)
\phi(x)=(i\hbar\partial_t+{c^2\hbar^2\over2{W}}\triangle)\psi(x)=0;
\end{equation}
$$
\psi_1 (x)={exp\lbrack-{i\over\hbar}{({\beta^2\over2}W{t}-\rm \bf
{P.x})}\rbrack}=
exp\lbrack-i{mv^2\over2\hbar}(t-{{\bf {s.x}}\over{v}/2})\rbrack;
$$
\begin{equation}
\label{f21}
\psi_2 (x)=exp\lbrack-{i\over\hbar}(Wt-\sqrt2{{\rm \bf P.x}\over\beta})\rbrack=
exp\lbrack-i{mc^2\over\hbar}(t-{{\bf s.x}\over c/\sqrt2})\rbrack.
\end{equation}
Generator of space-time transformations and its commutational properties:
\begin{equation}
\label{f22}
M_{01}=x_0\partial_1-x_1\partial_0, \ [L_S [L_S,M_{01}]]=0.
\end{equation}
Conditions of transfer of equation  (\ref{f20}) into itself:
\begin{equation}
\label{f23}
\{ i\hbar(\partial_t+V\partial_x)+{c^2 \hbar^2 (1-V^2/c^2) 
\over 2W(1-{\bf V.v}/c^2)}\lbrack{(\partial_x+V\partial_t/c^2)^2
\over(1-V^2/c^2)}+\partial_{yy}+\partial_{zz} \rbrack \} \Psi \psi =0; \
L_S \psi =0.
\end{equation}
Weight functions:
$$
\Psi_{11}=exp\{ -i{W\over2\hbar(1 - \beta^2)}[ ({\beta'_v}^2 -2\beta^2 - 
{\beta_v}^2(1 - \beta^2) - \beta \beta_x ({\beta'_v}^2 - 2))t - ({\beta'_v}^2 
- 2)(\beta - \beta^2 \beta_x ){x\over c}]\};
$$
$$
\begin{array}{l}
\displaystyle
\hspace{-15mm}\Psi_{22}=exp\{ -i{W\over2\hbar(1-\beta^2)}[(1-
{\sqrt2\over \beta'_v})(\beta^2-\beta \beta_x)t+ \\
\displaystyle
\hspace{-15mm}((1-{\sqrt2\over \beta'_v})(\beta^2 \beta_x-\beta)+
\sqrt2\beta_x({1\over \beta_v}-{1\over \beta'_v})){x\over c}+
\sqrt2(1-\beta^2)({1\over \beta_v}-{1\over \beta'_v})(\beta_y {y\over c}+
\beta_z {z\over c})]\};
\end{array}
$$
\begin{equation}
\label{f24}
\begin{array}{l}
\hspace{-82mm}\Psi_{12}=\Psi_{11} \psi_1 /\psi_2; \ \Psi_{21}=\Psi_{22} 
\psi_2/\psi_1.
\end{array}
\end{equation}
Transformational properties of solutions (\ref{f21}):
\begin{equation}
\label{f25}
\psi'_1 (x')=\Psi_{11} (x)\psi_1 (x)=\Psi_{12} (x)\psi_2 (x); \ 
\psi'_2 (x')=\Psi_{21} (x)\psi_1 (x)=\Psi_{22} (x)\psi_2 (x).
\end{equation}
\begin{sloppypar}
\par In   the   non-relativistic   approximation   the   equation
(\ref{f20}) and his solution $\psi_1 (x)$ and
the weight function $\Psi_{11} (x)$
obtain the known view  $L_S\psi(x)=0 \to (i\hbar\partial_t+\hbar^2 \triangle
/2m_0 )\psi(x)=0$; \
$\psi_1 (x) \to exp\lbrack-i(m_0 v^2/2\hbar)(t-{\bf s.x}/(v/2))=
exp\lbrack-i({\rm E}t-{\bf x.p})/\hbar \rbrack$; \
$\Psi_{11} (x) \to exp\lbrack-i(-{\sl E}t+x{\sl P})/\hbar \rbrack$ 
\cite{Fus90}, 
where $\rm E=m_0{\bf v}^2/2$;
${\sl E}=m_0 V^2/2$; ${\bf p}=m_0{\bf v}$; ${\sl P}=m_0 V$, $W=mc^2$, 
${\bf P}=m{\bf v}$, ${\bf s}={\bf v}/v$, ${\bf n}={\bf c}/c$;
$V$ is the velocity of inertial reference $K'$ relative to $K$;  
$\beta =V/c$; $\lambda =c'/c=(1 - 2 \beta n_x + \beta ^2)^{1/2}$; 
$\beta_x =v_x/c$, $\beta_y=v_y/c$, $\beta_z=v_z/c$, 
$\beta_v =v/c$; ${\bf v}=(v_x, v_y, v_z)$ is the speed of a particle; $c$ is
the speed of light;
$m=m_0/(1 - \beta^2)^{1/2}$ is the relativistic mass, $m_0$ is the rest mass
of a particle; ${\beta'_v}^2 =[\beta^2 (1 - \beta_v^2) + \beta_v^2 -
2\beta \beta_x + \beta^2 \beta_x^2 ]/(1 - \beta \beta_x)^2$. 
\par The second solution (\ref{f21}) is $\psi_2 (x) \to exp\lbrack-
i(m_0 c^2/\hbar)(t-{\bf s.x}/(c/\sqrt2)$. 
It is the new solution of the  non-relativistic Schr\"odinger equation 
\cite{Kot94}.
\end{sloppypar}
\subsection{The Galilei symmetry of Maxwell equations, p=2}
\begin{equation}
\label{f26}
\nabla .\ {\bf E} =0; \ {}
\nabla \times {\bf H} - {1\over c}\partial_t
{\bf E}=0; \ {} 
\nabla .\ {\bf H} = 0; \ {}
\nabla \times {\bf E} + {1\over c}\partial_t
{\bf H} = 0;
\end{equation}
\begin{equation}
\label{f27}
({\bf E}, {\bf H})=({\bf l},{\bf m}) exp(-ik.x)=
({\bf l},{\bf m}) \omega (t - {\bf n.x }/c),
\end{equation}
where ${\bf l}$, ${\bf m}$ are the vectors of polarization. 
We find the field transformation law as
\begin{equation}
\label{f28}
\begin{array}{ll}
{E_1}'=\Phi_D (x) E_1;  & {H_1}'=\Phi_D (x) H_1; \\
{E_2}'=\Phi_D (x) k(E_2 + h_{23}H_3); & 
\displaystyle {H_2}'=\Phi_D (x) k(H_2 + e_{23}E_3); \\
{E_3}'=\Phi_D (x) k(E_3 + h_{32}H_2); & 
\displaystyle {H_3}'=\Phi_D (x) k(H_3 + e_{32}E_2).
\end{array}
\end{equation} 
Here $\Phi_D$ is the weight function (\ref{f18}), $k$, $e_{23}$, $e_{32}$, 
$h_{23}$, $h_{32}$ are parametrs of transformations.
Bear in mind the expressions (\ref{f28}) and replacing the variables 
in Eq. (\ref{f26}) we
receive the system of  enganging equations analogous to Sys. (\ref{f17}) and
(\ref{f23}). Insertion the solution (\ref{f27}) and weight function (\ref{f18})
in this system  leads to superdefined system of algebraic equations for 
determination of parameters $k$, $e_{23}$, \ldots . The system has solutions:
\begin{equation}
\label{f29}
k={n_x(\beta-n_x)+\lambda\over {1-{n_x}^2}}; \
e_{23}={n_x (\lambda-1)+\beta\over {n_x (\beta-n_x)+\lambda}}; \
h_{23}=-{n_x (\lambda-1)+\beta\over {n_x (\beta-n_x)+\lambda}},
\end{equation}
where $e_{23}=-e_{32}=h_{32}$, $h_{23}=-h_{32}=e_{32}$. 
The parameters and 
weight fuction have the following transformational properties
because of the
Galilei addition theorem of velocities $\beta^{''}=\beta+\lambda \beta'$ and
transformation law of guiding cosines $n'_x=(n_x-\beta)/\lambda$, $n'_y=
n_y/\lambda$, $n'_z=n_z/\lambda$ 
\begin{equation}
\label{f30}
\Phi^{''}=\Phi' \Phi; \ d^{''}=(d'+d)/(1+d'd); \ k^{''}=k'k(1+d'd),
\end{equation}
where $d$=($e_{23}$, $e_{32}$, $h_{23}$, $h_{32}$), $\beta^{''}=V^{''}/c$, 
$\beta'=V'/c'$, $\beta=V/c$, $\lambda=c'/c$.
For comparison, in relativistic theory with $\Phi=1$, $k=1/(1-\beta^2)^{1/2}$,
$e_{23}=\beta$, $h_{23}=-\beta$
the relations (\ref{f30}) are valid
because of the relativistic addition theorem of velocities $\beta^{''}=
(\beta'+\beta)/(1+\beta'\beta)$.
Transfomations 
of fields (\ref{f28})
hold invariance of the forms ${\bf E}'{\bf H}'=k^2\Phi^2{\bf EH}=0$, 
\ $E'^2-H'^2=k^2\Phi^2 (E^2-H^2)=0$.
When parameter $\beta \to 0$,
the  weight  function $\Phi_D(x)\approx 1$,
the  transformations of the electric and magnetic fields
have a limit 
${\bf E}\approx {\bf E}+\beta\times {\bf H}$, \
${\bf H}\approx {\bf H}-\beta\times {\bf E}$ coinsident with 
the known non-relativistic one. It is the common limit for the relativistic 
and Galilei transformations of electromagnetic field.
This result is similar to the one, known as non-relativistic limit of the 
space-time transformatins $x'=x-Vt$, $y'=y$, $z'=z$, $t'=t$, $c'=c$ in the
relativistic theory. Being neither Lorentz transformations nor Galilei ones,
these transformations are the same limit for both Lorentz and  
Galilei space and time transformations indeed.
\section{Conclusion}
In summary it is possible to state, that the concept of symmetry is 
conventional. Dividing  equations into the
relativistic and the Galilei-invariant equations makes sense only in the
case, when $p=1$. In more general case, when
$p\geq 1$, equations have cumulative symmetrical properties
complying with the principles of relativity in the relativistic,
in the Galilei, as well as in the other versions.
In particular, Poincar\'e group ${\rm P}_{10}$ is the classical group of
symmetry of D'Alembert and Maxwell equations with $p=1$ and reflects the
property of relativistic invariance of these equations. The Galilei
group ${\rm G}_{10}$ is the generalized group of symmetry of D'Alembert
and Maxwell equations with $p=2$ and reflects the property
of invariance of these equations relative to the space and time 
transformations of the classical physics.
\par Analogous situation takes place in the case of Schr\"odinger equation
with the difference that the Galilei group is not the classical one but
the generalized group of symmetry of the equation with $p=1$ and the
Poincar\'e group is the generalized group of symmetry with $p=2$.
\par Both groups are the subgroups of the 20-dimentional group
of inhomogeneous linear space and time transformations IGL(4,R) in the space
$R^4{(x)}$. This group is the maximal linear group symmetry \cite{Kot94} of
the discussed equations in view of relations
\begin{equation}
\label{f31}
[\Box, P_a]=[L^r_s,  P_a]  =  0{;}\  [\Box  [\Box,  G_{ab}]]=[L^r_s  [L^r_s,
G_{ab}]] = 0,
\end{equation}
where $P_a=\partial _a$, $G_{ab}=x^a \partial _b$.
Owing to this relations, the Galilei symmetry of the D'Alembert and of the 
free Maxwell equations and relativistic symmetry of the free Schr\"odinger 
equation (the new symmetries of these equations) are not exotic
but their natural properties because the generators of corresponding
groups may be composed from the generators of Lie algebra of IGL(4,R) group.


\begin{thebibliography}{99}
\bibitem{Ovs78} L. V. Ovsyannikov, Group analysis of differential
equations. Nauka, Moskow, 1978. 
\bibitem{Fus90} V. I. Fushchich, A. G. Nikitin, The Symmetry of the Quantum
Mechanics Equations. Nauka, Moscow, 1990, p. 9-10, 159-170, 280.
\bibitem{Zhe86} D. P. Zhelobenko, in: {\it Group Theoretical Methods in
Physics, v. 2}, Proceedings of the Third Yurmala Seminar, editted by
M. A. Markov, V. I. Man'ko, V. V. Dodonov. Nauka, Moscow, 1986, p. 5-21.
\bibitem{Shi94} D. V. Shirkov, UMN (Russia), 1994, v. 49, N 5(299), p. 145-164.
\bibitem{Vor86} E. M. Vorob'ev, DAN (USSR), 1986, v. 287, N 3, p. 536-539.
\bibitem{Olv86} P. J. Olver, P. Rosenau, Phys. Lett., 1986, v. 114A, 
N 3, p. 107-112.
\bibitem{Mal79} I. A. Malkin, V. I. Man'ko, The Dynamic Symmetry and
Coherent States of Quantum Systems. Nauka, Moscow, 1979, p. 17.
\bibitem{Lez86} A. N. Leznov, V. I. Man'ko, M. V. Savel'ev, Soliton
Solutions of Nonlinear Equations and the Theory of Group
Representations. Proceedings of FIAN, v. 165. Nauka, Moscow, 1986, p. 75.
\bibitem{Kot83} G. A. Kotel'nikov, in: {\it Group Theoretical Methods in
Physics, v. 1}, Proceedings of the Second Zvenigorod Seminar, edited by
M. A. Markov, V. I. Man'ko, A. E. Shabad. Harwood Academic Publichers,
Chur, London, Paris, New York, 1985, p. 507-516.
\bibitem{Kot86} G. A. Kotel'nikov, in: {\it Group Theoretical Methods in
Physics, v. 2}, Proceedings of the Third Yurmala Seminar, editted by
M. A. Markov, V. I. Man'ko, V. V. Dodonov. VNU Science, Utrecht,
The Netherlands, 1986, p. 95-109.
\bibitem{Kot94} G. A. Kotel'nikov, Preprint IAE-5778/1, M., 1994, 21 p.

\end{thebibliography}
\end{document}